\documentclass[11pt,a4paper]{article}
\usepackage{amsmath}
\usepackage{amsthm}
\usepackage{amssymb}
\usepackage{graphicx}
\usepackage[cp1251]{inputenc}
\usepackage[T2A]{fontenc}

\usepackage{graphicx}
\usepackage[all]{xy}

\newcommand{\beq}[1]{\begin{equation}\label{#1}}
\newcommand{\eq}{\end{equation}}

\begin{document}
\setcounter{footnote}{1}
\def\thefootnote{\fnsymbol{footnote}}

\vspace{1cm}

\begin{flushright}
ITEP-TH-42/17
\end{flushright}

\begin{center}
\textsc{
Bruhat order in the Toda system on $\mathfrak{so}(2,4)$:\\ an example of non-split real form}\\

\vspace{.5cm}
\ \\
Yu.B. Chernyakov\footnote{Institute for Theoretical and Experimental Physics, Bolshaya Cheremushkinskaya, 25,
117218 Moscow, Russia.}$^{,}$\footnote{Joint Institute for Nuclear Research, Bogoliubov Laboratory of Theoretical Physics, 141980 Dubna, Moscow region, Russia.}, chernyakov@itep.ru\\
G.I Sharygin\footnotemark[2]$^{,}$\footnotemark[3]$^{,}$\footnote{Lomonosov
Moscow State University, Faculty of Mechanics and Mathematics, GSP-1, 1 Leninskiye Gory, Main Building, 119991 Moscow, Russia.}, sharygin@itep.ru\\
A.S. Sorin\footnotemark[3],$^{,}$\footnote{National Research Nuclear University MEPhI
(Moscow Engineering Physics Institute),
Kashirskoye shosse 31, 115409 Moscow, Russia}$^{,}$\footnote{Dubna International University,
141980 Dubna (Moscow region), Russia.} sorin@theor.jinr.ru
\end{center}

\begin{abstract}
In our previous papers (\cite{CSS14,CSS15,CSS17}) we described the structure of trajectories of the symmetric Toda system on normal real forms of various Lie algebras and showed that it was totally determined by the Hasse diagram of the Bruhat order on the corresponding Weil group. This note deals with the simplest non-split real Lie algebra, $\mathfrak{so}(2,4)$. It turns out, that the phase diagram in this case is also closely related with the Bruhat order of the relative Weyl group, but a bit more information is necessary to describe the dimensions of the trajectory spaces.
\end{abstract}

\section{Introduction}
Let us briefly recall the definitions of Toda system on a semisimple real Lie algebra and its properties; the principal source for this part is Helgason's and Vinberg-Onischik books \cite{Hel, VinOn} and the papers \cite{deMPe, BBR, BG}.
\subsection{Lie algebras, Weyl groups and Bruhat order}
So, let $\mathfrak g_\mathbb C$ be a semisimple complex Lie algebra, and $\mathfrak g$ its real form; $G$ be the corresponding Lie group. We fix a Cartan involution of $\mathfrak g,\ \theta:\mathfrak g\to \mathfrak g$ and the corresponding Cartan decomposition:
\begin{equation}
\label{eq:Card}
\mathfrak g=\mathfrak k\oplus \mathfrak p,
\end{equation}
where $\mathfrak k$, the $+1$ eigenspace of $\theta$, is a maximal compact subalgebra of $\mathfrak g$ and $\mathfrak p$, the $-1$ eigenspace of $\theta$, is the linear complement to $\mathfrak k$, such that
\[
{}[\mathfrak k,\mathfrak p]\subset \mathfrak p,\ [\mathfrak p,\mathfrak p]\subset \mathfrak k.
\]
It follows that all subalgebras of $\mathfrak p$ are commutative.

Let $\mathfrak a$ be a maximal (with respect to inclusions) subalgebra of $\mathfrak p$. One says that the real form $\mathfrak g$ is \textit{normal}, or \textit{split} if $\mathfrak a$ commutes with no nontrivial element in $\mathfrak k$ (equivalently, if $\mathfrak a=\mathfrak h$, the maximal commutative subalgebra in $\mathfrak g$). Since all Cartan decompositions of $\mathfrak g$ are equivalent, this notion depends only on $\mathfrak g$. Otherwise, the real form $\mathfrak g$ is called \textit{non-split}.

We fix the maximal commutative subalgebra $\mathfrak a$ in $\mathfrak p$ and find the root decomposition of $\mathfrak g$ (in this case roots are considered as linear functionals on $\mathfrak a$, see \cite{Hel}):
\[
\mathfrak g=\mathfrak a\oplus\bigoplus_{\alpha\in\Phi_+}(\mathfrak g_\alpha\oplus\mathfrak g_{-\alpha}),
\]
where $\Phi_+$ is the set of positive roots with respect to a fixed Weyl chamber and $\mathfrak g_\alpha$ is the root space, corresponding to $\alpha$; let $\Delta$ denote corresponding set of simple roots. In this case we obtain the Iwasawa decomposition:
\[
\mathfrak g=\mathfrak k\oplus\mathfrak a\oplus\mathfrak n,
\]
where $\mathfrak n=\bigoplus_{\alpha\in\Phi_+}\mathfrak g_\alpha$. In the case of normal real form $\mathfrak g_\alpha$ are one-dimensional for all $\alpha$, and in all cases there is a canonical identification of $\mathfrak g_\alpha$ and $\mathfrak g_{-\alpha}$ given by the application of Cartan involution $\theta$. We choose the basis vectors $e_\alpha$ in $\mathfrak g_\alpha$ and let $e_{-\alpha}$ be the corresponding basis in $\mathfrak g_{-\alpha}$ (in general, there can be more than one vector $e_\alpha$, by a slight abuse of notation). In this notation we can decompose $\mathfrak k$ and $\mathfrak p$ as
\[
\begin{aligned}
\mathfrak k&=\bigoplus_{\alpha\in\Phi_+}\mathbb R(e_\alpha-e_{-\alpha}),\\
\mathfrak p&=\mathfrak a\oplus\bigoplus_{\alpha\in\Phi_+}\mathbb R(e_\alpha+e_{-\alpha}).
\end{aligned}
\]
Using this notation we define the \textit{Toda symmetrization map} $M:\mathfrak p\to\mathfrak k$ by equation:
\[
M\left(a+\sum_{\alpha\in\Phi_+}b_\alpha(e_\alpha+e_{-\alpha})\right)=\sum_{\alpha\in\Phi_+}b_\alpha(e_\alpha-e_{-\alpha}).
\]
Clearly, it does not depend on the choice of the basis $e_\alpha$.

Let now $K$ be the maximal compact subgroup of $G$, corresponding to $\mathfrak k$. By \eqref{eq:Card} it follows that the adjoint representation of $K$ on $\mathfrak g$ preserves $\mathfrak p$. Let $Z_K(\mathfrak a),\,N_K(\mathfrak a)$ be the centralizer and normalizer of $\mathfrak a$ in $K$ with respect to this action; clearly, $Z_K(\mathfrak a)$ is a normal subgroup of $N_K(\mathfrak a)$. In the case when the real form $\mathfrak g$ is normal, both groups $Z_K(\mathfrak a)$ and $N_K(\mathfrak a)$ are discrete, but in non-split case these groups have a common Lie algebra. In all cases the \textit{Weyl group of the real form $\mathfrak g$} (sometimes called restricted Weyl group) is equal to the quotient:
\[
W(\mathfrak g,\mathfrak k)=N_K(\mathfrak a)/Z_K(\mathfrak a).
\]
The group $W(\mathfrak g,\mathfrak k)$ is discrete and is naturally embedded into the \textit{flag space of $K$}, $Fl(K)=K/Z_K(\mathfrak a)$; it is equipped with a natural exact representation on $\mathfrak a$. In terms of this representation one can show that $W(\mathfrak g,\mathfrak k)$ is always generated by a finite set of reflections: reflections with respect to the hyperplanes determined by simple roots. Let $w_1,\dots,w_k$ be these reflections, here $k$ is the number of simple roots and $w_i=w_{\alpha_i},\ \alpha_i\in\Delta$.

One defines the \textit{length} $l(w)$ of an element $w$ in $W(\mathfrak g,\mathfrak k)$ as the number $n$ of generators in the minimal word $w=w_{i_1}w_{i_2}\dots w_{i_n}$, representing $w$. Similarly, one defines the \textit{(strong) Bruhat order} on $W(\mathfrak g,\mathfrak k)$ as the partial order, determined by the following elementary relation:
\[
u\prec_1 v,\ \mbox{iff}\ v=w_\alpha u,\ \mbox{for some $\alpha\in\Delta$, and}\ l(v)=l(u)+1.
\]
In other words, the Bruhat order on $W(\mathfrak g,\mathfrak k)$ is the minimal partial order $\prec$, generated by relations $\prec_1$ and the transitivity condition.

\subsection{Toda system}
Observe, that the Iwasawa decomposition and Killing form on $\mathfrak g$ allows one identify $\mathfrak p$ with the dual space of the maximal solvable subalgebra in $\mathfrak g$, the \textit{Borel subalgebra} $\mathfrak b=\mathfrak a\oplus\mathfrak n$, and thus there is a natural Poisson structure on $\mathfrak p$, pulled from $\mathfrak b^*$. \textit{Symmetric Toda system} on $\mathfrak p$ is the Hamiltonian system determined by this Poisson structure and the Hamiltonian function $H(L)=Tr(ad(L)^2),\ L\in\mathfrak p$ (the adjoint representation is taken with respect to $\mathfrak g$), $L$ -- matrix of the Lax operator.

There are many methods to solve and describe the trajectories of this system. We shall combine two of them. First of all, it turns out that the Toda dynamics on $\mathfrak p$ can be described in the terms of the adjoint action and a gradient system on $K$. Namely, fix $L\in\mathfrak p$ and consider the following vector field on $K$:
\[
\mathcal T_L(k)=dR_k(M(Ad_k(L))),\ k\in K,
\]
where $R_k$ is the right translation by $k\in K$. Then \textit{the trajectory of Toda system with that initiates at $L\in\mathfrak p$ is given by the formula: $L(t)=Ad_{k(t)}(L)$, where $k(t)$ is the trajectory of $\mathcal T_L$ with $k(0)=e$.}

Thus we are brought to the study of the properties of the vector fields $\mathcal T_L$ on $K$ for various $L\in\mathfrak p$. It follows from the description just given, that the trajectories never leave the adjoint orbits of $K$ in $\mathfrak p$, thus for the rest of the paper we fix such an orbit; since all the orbits of $K$ pass through $\mathfrak a\subseteq\mathfrak p$, we can always assume that $L=\Lambda\in\mathfrak a$.

Moreover, we shall assume that $\Lambda$ is a regular element of $\mathfrak a$, in particular its centralizer in $K$ coincides with $Z_K(\mathfrak a)$. So the field $\mathcal T_\Lambda$ is $Z_K(\mathfrak a)$-invariant and one can regard it as a field on the flag space $Fl(K)$.

In this setting it is clear that the points of $W(\mathfrak g,\mathfrak k)\subset Fl(K)$ and only they are singular for $\mathcal T_\Lambda$. Moreover, one can show (see \cite{deMPe,BG}) that this field is in effect equal to the gradient of a smooth function $F(k)$ on $Fl(K)$, which in fact is a Morse function. Below, however, we shall usually work on the group $K$, rather than on the flag space; in this case the set of singularities of the field $\mathcal T_\Lambda$ coincides with $N_K(\mathfrak a)$ and this field is gradient for a Morse-Bott function. One should only keep in mind that the gradient both times is taken with respect to an invariant Riemann structure, but not the one, given by Killing form (see \cite{deMPe,BG} for details; below we shall give an explicit formula for this function in the particular case we consider).

This property makes it possible to use elements of the Morse theory for studying the Toda system on $K$. Recall, that for every singular point $k$ of $F$, the gradient vector field in a neighbourhood of $k$ is determined by the Hessian $d^2F$ of $F$ at $k$: the incoming trajectories (the \textit{unstable submanifold}) of $k$ are tangent to the subspace spanned by the coordinates with negative squares in the canonical form of $d^2F$ in $T_kK$ and the exiting trajectories (\textit{stable submanifold}) are tangent to the space of variables whose squares enter this expression with positive coefficients. The index of $k$ is the number of negative squares in the Hessian.

On the other hand, if we embed the real Lie algebra $\mathfrak g$ into a bigger one $I:\mathfrak g\hookrightarrow \mathfrak g_1$ so that the Cartan involutions commute with the embedding homomorphism, then it is clear that the Toda system on $\mathfrak p$ is equal to a restriction of the corresponding system on $\mathfrak p_1$.

Similarly, if we consider the induced embedding of the Lie groups $K\hookrightarrow K_1$ then the field $\mathcal T_\Lambda$ on $K$ is equal to the restriction of the field $\mathcal T^1_\Lambda$ on $K_1$, for any element $\Lambda\in\mathfrak p\subseteq\mathfrak p_1$ (we even can assume that the maximal subalgebra $\mathfrak a$ of $\mathfrak p$ lies inside the maximal subalgebra $\mathfrak a_1$ of $\mathfrak p_1$, since every commutative subalgebra can be embedded into a maximal). This, however, does not mean that the function $F$ is equal to the restriction of $F_1$.

We shall use this observation in the following context: suppose, we chose an embedding of $\mathfrak g$ into $\mathfrak{sl}(n)$ so that the Cartan involution is preserved. In this case $\mathfrak k$ is embedded into $\mathfrak{so}(n)$, $\mathfrak p$ goes into $Sym_0(n,\mathbb R)$ (the space of traceless real symmetric matrices), $\mathfrak a\subseteq diag_0(n,\mathbb R)$ (real diagonal traceless matrices) and the dynamics on $K$ is equal to the restriction of the dynamics on $SO(n,\mathbb R)$. So one can find invariant subspaces in $K$ as intersections of this group and invariant subspaces of the full symmetric Toda flow on $SO(n,\mathbb R)$.

This case is well studied and a vast family of invariant subvarieties in $SO(n,\mathbb R)$ called the \textit{minor surfaces} is known (see \cite{CS,FS,CSS14} for details): they are given by the solution of polynomial equations of the form
\[
M_{i_1\dots i_k}(\Psi)=0,\ \Psi\in SO(n,\mathbb R).
\]
Here $M_{i_1\dots i_k}(\Psi)$ is the determinant of submatrix in $\Psi$, spanned by the first $k$ rows and the columns $i_1,\,i_2,\dots,i_k$. Using the fact that the system on $K$ is gradient and that the local picture in a neighbourhood of critical points is determined by the Hessian of the function, and choosing carefully minor surfaces, one can find the phase portrait of the system.

In the following section we shall apply this method to study the system on $SO(2,4)$, the group of matrices, that preserve the quadratic form
\begin{equation}
\label{eq:qua}
q(x)=x_1^2+x_2^2-x_3^2-x_4^2-x_5^2-x_6^2
\end{equation}
on $\mathbb R^6$. In our previous papers (\cite{CSS14,CSS15,CSS17}) we used this approach to show the phase portrait of Toda flow on normal real forms of semisimple algebras. It turned out that in all these situations it coincides with the Hasse diagram of the Bruhat order on $W(\mathfrak g,\mathfrak k)$ (i.e. with the oriented graph, whose edges correspond to the pairs of elements $u\prec v$ in the Weyl group).

The case we consider here is the simplest non-split real form of a Lie algebra. In this paper we shall show that the same statement stays true in this case. However, here the dimensions of the subspaces, spanned by the trajectories, connecting two points, are determined not just by this combinatoric data, but rather by the indices of the Morse function in the points (see below). This is a new phenomenon and it is intriguing to find a general explanation of this structure. We shall postpone it to a forthcoming paper.

\section{The Toda system on $SO(2,4)$}
The group $SO(2,4)$ of matrices, preserving the quadratic form \eqref{eq:qua}; this is noncompact real Lie group. Its Lie algebra $\mathfrak{so}(2,4)$ consists of the matrices of the form (see \cite{Hel,VinOn}):
\[
\begin{pmatrix}C & X\\ X^T &B\end{pmatrix},
\]
where $C\in Mat_2(\mathbb R), B\in Mat_4(\mathbb R), X\in Mat_{2\times 4}(\mathbb R)$ ,and $C^T=-C,\ B^T=-B$. In this case
\[
\begin{aligned}
\mathfrak k&=\mathfrak{so}(2)\oplus\mathfrak{so}(4)\\
                  &=\left\{U=\left(
\begin{array}{cccccc}
 0 & \varphi & 0 & 0 & 0 & 0 \\
 -\varphi & 0 & 0 & 0 & 0 & 0 \\
 0 & 0 & 0 & \theta_{12} & \theta_{13} & \theta_{14} \\
 0 & 0 & -\theta_{12} & 0 & \theta_{23} & \theta_{24} \\
 0 & 0 & -\theta_{13} & -\theta_{23} & 0 & \theta_{34} \\
 0 & 0 & -\theta_{14} & -\theta_{24} & -\theta_{34} & 0
\end{array}
\right)\right\}\\
\mathfrak p&=\left\{V=\left(
\begin{array}{cccccc}
 0 & 0 & v_{11} & v_{12} & v_{13} & v_{14} \\
 0 & 0 & v_{21} & v_{22} & v_{23} & v_{24} \\
 v_{11} &  v_{21} & 0 & 0 & 0 & 0 \\
 v_{12} &  v_{22} & 0 & 0 & 0 & 0 \\
 v_{13} &  v_{23} & 0 & 0 & 0 & 0\\
 v_{14} &  v_{24} & 0 & 0 & 0 & 0
\end{array}
\right)\right\}\\
\mathfrak a&=\left\{W=\left(
\begin{array}{cccccc}
 0 & 0 & a & 0 & 0 & 0 \\
 0 & 0 & 0 & b & 0 & 0 \\
 a & 0 & 0 & 0 & 0 & 0 \\
 0 & b & 0 & 0 & 0 & 0 \\
 0 & 0 & 0 & 0 & 0 & 0 \\
 0 & 0 & 0 & 0 & 0 & 0
\end{array}
\right)\right\}.
\end{aligned}
\]
One can also describe the root spaces, normalizator and centralizer of $\mathfrak a$ and Weyl group in these terms: the root system and Weyl group is isomorphic to those of $C_2$ class, but the root spaces are not always $1$-dimensional any more. However, as one sees, this choice of representation does not verify the conditions, we need: $\mathfrak a$ does not fall into the space of diagonal matrices, so we shall not do this now.

Instead we shall conjugate this representation by the orthogonal matrix
\[
A=
\left(
\begin{array}{cccccc}
 \frac{1}{\sqrt{2}} & 0 & \frac{1}{\sqrt{2}} & 0 & 0 & 0 \\
 0 & \frac{1}{\sqrt{2}} & 0 & \frac{1}{\sqrt{2}} & 0 & 0\\
0 & 0 & 0 & 0 & 0 & 1 \\
 0 & 0 & 0 & 0 & 1 & 0 \\
 0 & -\frac{1}{\sqrt{2}} & 0 & \frac{1}{\sqrt{2}} & 0 & 0 \\
 -\frac{1}{\sqrt{2}} & 0 & \frac{1}{\sqrt{2}} & 0 & 0 & 0
 \end{array}
\right),
\]
which will send $\mathfrak a$ into
\begin{equation}
\label{eq:ab}
A\cdot \mathfrak a \cdot A^T=\left\{W'=\left(
\begin{array}{cccccc}
 a & 0 & 0 & 0 & 0 & 0 \\
 0 & b & 0 & 0 & 0 & 0 \\
 0 & 0 & 0 & 0 & 0 & 0 \\
 0 & 0 & 0 & 0 & 0 & 0 \\
 0 & 0 & 0 & 0 & -b & 0 \\
 0 & 0 & 0 & 0 & 0 & -a
\end{array}
\right)
\right\}
\end{equation}
and preserve the spaces of symmetric and antisymmetric matrices; clearly, there are more than one orthogonal matrix $A$ with this property; our choice is partly dictated by our intention to make our formulas look like those in \cite{FS}. Now, after this change of coordinates, all the conditions on the matrices, representing $\mathfrak k,\,\mathfrak p$ and $\mathfrak a$ hold; in particular the maximal compact subgroup of $SO(2,4)$ is sent to the following subgroup $A\cdot (SO(2,\mathbb R)\times SO(4,\mathbb R))\cdot A^T\subset SO(6,\mathbb R)$. As we explained before, Toda system on the maximal compact subgroup is now restricted from the group $SO(6,\mathbb R)$ via the embedding explained above, and the minor surfaces in the $SO(6,\mathbb R)$ generate invariant subspaces of the Toda flow.

Let us now describe the Morse function of the Toda field on $SO(2,4)$, (see \cite{deMPe,BG} for details): for a regular element $\Lambda\in\mathfrak a$, the field $\mathcal T_\Lambda$ is equal to the gradient of the following function $F_\Lambda(\Psi),\ \Psi\in K$
\[
F_\Lambda(\Psi)=Tr(ad(Ad_\Psi(\Lambda))ad(N)),
\]
where $N\in\mathfrak a$ is a special element, which one finds from the system of equations:
\[
\alpha\left(\sum_{\gamma\in\Delta}x_\gamma h_\gamma\right)=-1,\ \mbox{for all }\alpha\in\Delta.
\]
Here $h_\gamma$ is the standard basis of $\mathfrak a$, given by coroots. This system is solvable, since its matrix of coefficients (which is the Cartan matrix of the roots) is nondegenerate.

In the case of the matrix representation of $SO(2,4)$ that we use here, one has
\[
h_1=\left(
\begin{array}{cccccc}
 0 & 0 & 0 & 0 & 0 & 0 \\
 0 & 1 & 0 & 0 & 0 & 0 \\
 0 & 0 & 0 & 0 & 0 & 0 \\
 0 & 0 & 0 & 0 & 0 & 0 \\
 0 & 0 & 0 & 0 & -1 & 0 \\
 0 & 0 & 0 & 0 & 0 & 0
\end{array}
\right),\ h_2=\left(
\begin{array}{cccccc}
 1 & 0 & 0 & 0 & 0 & 0 \\
 0 & -1 & 0 & 0 & 0 & 0 \\
 0 & 0 & 0 & 0 & 0 & 0 \\
 0 & 0 & 0 & 0 & 0 & 0 \\
 0 & 0 & 0 & 0 & 1 & 0 \\
 0 & 0 & 0 & 0 & 0 & -1
\end{array}
\right)
\]
and the Cartan matrix is $\left(
\begin{array}{cc}
 2 & -1 \\
 -2 & 2
\end{array}
\right)$, whence $x_1 = -3/2,\, x_2 = -2$. Finally,
\[
F_\Lambda(\Psi)=Tr(\Psi\Lambda\Psi^TN),
\]
where the matrix $N$ is equal to $x_1h_1+x_2h_2$, so
\[
N=\left(
\begin{array}{cccccc}
 -2 & 0 & 0 & 0 & 0 & 0 \\
 0 & 1/2 & 0 & 0 & 0 & 0 \\
 0 & 0 & 0 & 0 & 0 & 0 \\
 0 & 0 & 0 & 0 & 0 & 0 \\
 0 & 0 & 0 & 0 & -1/2 & 0 \\
 0 & 0 & 0 & 0 & 0 & 2
\end{array}
\right).
\]
This matrix representation is very useful when we need to compute the Hessians of $F_\Lambda$ at its critical points. So we have now to describe the critical points of this system: as we know, this set is equal to the normalizer of $\mathfrak a$ in $K$: in the matrix representation we use here, this subgroup is generated by the centralizer $Z_K(\mathfrak a)$ and the following set of matrices
\[
\begin{array}{c}
w_1=
\left(
\begin{array}{cccccc}
 1 & 0 & 0 & 0 & 0 & 0 \\
 0 & 1 & 0 & 0 & 0 & 0 \\
 0 & 0 & 1 & 0 & 0 & 0 \\
 0 & 0 & 0 & 1 & 0 & 0 \\
 0 & 0 & 0 & 0 & 1 & 0 \\
 0 & 0 & 0 & 0 & 0 & 1
\end{array}
\right), \

w_2=
\left(
\begin{array}{cccccc}
 0 & 0 & 0 & 0 & 0 & -1 \\
 0 & 1 & 0 & 0 & 0 & 0 \\
 0 & 0 & 1 & 0 & 0 & 0 \\
 0 & 0 & 0 & -1 & 0 & 0 \\
 0 & 0 & 0 & 0 & 1 & 0 \\
 -1 & 0 & 0 & 0 & 0 & 0
\end{array}
\right)\\,

\ \\

w_3=
\left(
\begin{array}{cccccc}
 1 & 0 & 0 & 0 & 0 & 0 \\
 0 & 0 & 0 & 0 & -1 & 0 \\
 0 & 0 & -1 & 0 & 0 & 0 \\
 0 & 0 & 0 & 1 & 0 & 0 \\
 0 & -1 & 0 & 0 & 0 & 0 \\
 0 & 0 & 0 & 0 & 0 & 1
\end{array}
\right), \

w_4=
\left(
\begin{array}{cccccc}
 0 & 0 & 0 & 0 & 0 & -1 \\
 0 & 0 & 0 & 0 & -1 & 0 \\
 0 & 0 & -1 & 0 & 0 & 0 \\
 0 & 0 & 0 & -1 & 0 & 0 \\
 0 & -1 & 0 & 0 & 0 & 0 \\
 -1 & 0 & 0 & 0 & 0 & 0
\end{array}
\right)\\,

\ \\

w_5=
\left(
\begin{array}{cccccc}
 0 & 1 & 0 & 0 & 0 & 0 \\
 1 & 0 & 0 & 0 & 0 & 0 \\
 0 & 0 & 1 & 0 & 0 & 0 \\
 0 & 0 & 0 & 1 & 0 & 0 \\
 0 & 0 & 0 & 0 & 0 & 1 \\
 0 & 0 & 0 & 0 & 1 & 0
\end{array}
\right), \

w_6=
\left(
\begin{array}{cccccc}
 0 & 0 & 0 & 0 & 1 & 0 \\
 0 & 0 & 0 & 0 & 0 & 1 \\
 0 & 0 & 1 & 0 & 0 & 0 \\
 0 & 0 & 0 & 1 & 0 & 0 \\
 1 & 0 & 0 & 0 & 0 & 0 \\
 0 & 1 & 0 & 0 & 0 & 0
\end{array}
\right)\\,

\ \\

w_7=
\left(
\begin{array}{cccccc}
 0 & 1 & 0 & 0 & 0 & 0 \\
 0 & 0 & 0 & 0 & 0 & 1 \\
 0 & 0 & 1 & 0 & 0 & 0 \\
 0 & 0 & 0 & -1 & 0 & 0 \\
 1 & 0 & 0 & 0 & 0 & 0 \\
 0 & 0 & 0 & 0 & 1 & 0
\end{array}
\right), \

w_8=
\left(
\begin{array}{cccccc}
 0 & 0 & 0 & 0 & 1 & 0 \\
 1 & 0 & 0 & 0 & 0 & 0 \\
 0 & 0 & 1 & 0 & 0 & 0 \\
 0 & 0 & 0 & -1 & 0 & 0 \\
 0 & 0 & 0 & 0 & 0 & 1 \\
 0 & 1 & 0 & 0 & 0 & 0
\end{array}
\right).

\end{array}
\]
The real Lie algebra $\mathfrak{so}(2,4)$ being non-split, the centralizer in its turn is equal to the semidirect product of the $1$-dimensional subgroup
\[
SO_\mathfrak a(2,\mathbb R)=\left\{\left(
\begin{array}{cccccc}
 1 & 0 & 0 & 0 & 0 & 0 \\
 0 & 1 & 0 & 0 & 0 & 0 \\
 0 & 0 & \cos (\theta ) & \sin (\theta ) & 0 & 0 \\
 0 & 0 & -\sin (\theta ) & \cos (\theta ) & 0 & 0 \\
 0 & 0 & 0 & 0 & 1 & 0 \\
 0 & 0 & 0 & 0 & 0 & 1
\end{array}
\right)
\right\},
\]
and the following discrete subgroup of order $4$
\[
\begin{array}{c}
n_{1}=\left(
\begin{array}{cccccc}
 1 & 0 & 0 & 0 & 0 & 0 \\
 0 & 1 & 0 & 0 & 0 & 0 \\
 0 & 0 & 1 & 0 & 0 & 0 \\
 0 & 0 & 0 & 1 & 0 & 0 \\
 0 & 0 & 0 & 0 & 1 & 0 \\
 0 & 0 & 0 & 0 & 0 & 1
\end{array}
\right), \ \ \

n_{2}=\left(
\begin{array}{cccccc}
 -1 & 0 & 0 & 0 & 0 & 0 \\
 0 & -1 & 0 & 0 & 0 & 0 \\
 0 & 0 & 1 & 0 & 0 & 0 \\
 0 & 0 & 0 & 1 & 0 & 0 \\
 0 & 0 & 0 & 0 & -1 & 0 \\
 0 & 0 & 0 & 0 & 0 & -1
\end{array}
\right),\\

\ \\

n_{3}=\left(
\begin{array}{cccccc}
 1 & 0 & 0 & 0 & 0 & 0 \\
 0 & 1 & 0 & 0 & 0 & 0 \\
 0 & 0 & -1 & 0 & 0 & 0 \\
 0 & 0 & 0 & -1 & 0 & 0 \\
 0 & 0 & 0 & 0 & 1 & 0 \\
 0 & 0 & 0 & 0 & 0 & 1
\end{array}
\right), \ \ \

n_{4}=\left(
\begin{array}{cccccc}
 -1 & 0 & 0 & 0 & 0 & 0 \\
 0 & -1 & 0 & 0 & 0 & 0 \\
 0 & 0 & -1 & 0 & 0 & 0 \\
 0 & 0 & 0 & -1 & 0 & 0 \\
 0 & 0 & 0 & 0 & -1 & 0 \\
 0 & 0 & 0 & 0 & 0 & -1
\end{array}
\right).
\end{array}
\]
This means that $\dim Z_K(\mathfrak a)=1$; in particular, that the effective dimension of the phase space of the system, equal to the dimension of the flag space $Fl(K)$, is equal to $\dim(SO(2,\mathbb R)\times SO(4,\mathbb R))-1=6$.

Now we can ``assemble'' all this data and obtain the final result: first, we fix in $\Lambda$ to be equal to the matrix $W'$ (see the formula \eqref{eq:ab}) with $a=1,\,b=2$ and compute the indices of the Hessians of $F_\Lambda$ at its critical points using the local coordinates, pulled to $Fl(K)$ from the unit of the group, where we identify it with $\mathfrak{so}(2)\oplus\mathfrak{so}(4)/\mathfrak{so}_\mathfrak a(2)$; we regard this space as the set of matrices of the form
\[
\Theta=\left(
\begin{array}{cccccc}
 0 & -\theta _3 & -\theta _4 & -\theta _5 & -\theta _6 & 0 \\
 \theta _3 & 0 & -\theta _1 & -\theta _2 & 0 & \theta _6 \\
 \theta _4 & \theta _1 & 0 & 0 & \theta _1 & \theta _4 \\
 \theta _5 & \theta _2 & 0 & 0 & \theta _2 & \theta _5 \\
 \theta _6 & 0 & -\theta _1 & -\theta _2 & 0 & \theta _3 \\
 0 & -\theta _6 & -\theta _4 & -\theta _5 & -\theta _3 & 0
\end{array}
\right).
\]
The same computations allow one describe the positive and negative eigenvalues of the Hessian at these points. Next, we consider the intersections of certain minor surfaces in $SO(6,\mathbb R)$ with our maximal compact subgroup and write down the list of critical points in the flag space, which fall into one or another minor surface. It turns out, that to our purposes it is enough to consider only the simplest minor surfaces, corresponding to the set of $1\times1$ minors, i.e. given by the equations of the form $\psi_{1i}=0$. We represent the results of these computations in the form of the following table:
\begin{equation*}
\footnotesize{
\begin{tabular}{|c|c|c|c|}
\hline\cline{1-0}
$$ & $$ & $$ & $$\\
$i$ & $\theta _1,\theta _2,\theta _3,\theta _4,\theta _5,\theta _6$ & Index & Minors, $\psi_{1j}$\\

$$ & $$ & $$ & $$\\
\hline\cline{1-0}
$$ & $$ & $$ & $$\\
$w_1$ & $-,-,-,+,+,+$ & $3$ & $\psi_{12},\psi_{13},\psi_{14},\psi_{15},\psi_{16}$\\

$$ & $$ & $$ & $$\\
\hline\cline{1-0}
$$ & $$ & $$ & $$\\
$w_2$ & $-,-,-,-,-,+$ & $5$ & $\psi_{11},\psi_{12},\psi_{13},\psi_{14},\psi_{15}$\\

$$ & $$ & $$ & $$\\
\hline\cline{1-0}
$$ & $$ & $$ & $$\\
$w_3$ & $+,+,+,+,+,-$ & $1$ & $\psi_{12},\psi_{13},\psi_{14},\psi_{15},\psi_{16}$\\

$$ & $$ & $$ & $$\\
\hline\cline{1-0}
$$ & $$ & $$ & $$\\
$w_4$ & $+,+,+,-,-,-$ & $3$ & $\psi_{11},\psi_{12},\psi_{13},\psi_{14},\psi_{15}$\\

$$ & $$ & $$ & $$\\
\hline\cline{1-0}
$$ & $$ & $$ & $$\\
$w_5$ & $-,-,+,+,+,+$ & $2$ & $\psi_{11},\psi_{13},\psi_{14},\psi_{15},\psi_{16}$\\

$$ & $$ & $$ & $$\\
\hline\cline{1-0}
$$ & $$ & $$ & $$\\
$w_6$ & $+,+,-,-,-,-$ & $4$ & $\psi_{11},\psi_{12},\psi_{13},\psi_{14},\psi_{16}$\\

$$ & $$ & $$ & $$\\
\hline\cline{1-0}
$$ & $$ & $$ & $$\\
$w_7$ & $+,+,+,+,+,+$ & $0$ & $\psi_{11},\psi_{13},\psi_{14},\psi_{15},\psi_{16}$\\

$$ & $$ & $$ & $$\\
\hline\cline{1-0}
$$ & $$ & $$ & $$\\
$w_8$ & $-,-,-,-,-,-$ & $6$ & $\psi_{11},\psi_{12},\psi_{13},\psi_{14},\psi_{16}$\\

$$ & $$ & $$ & $$\\
\hline
\end{tabular}
\
}
\end{equation*}

\ \\

It turns out that in the coordinates $\theta_i$ these Hessians are always diagonal, so in this table, the second column gives the list of signs of the squares of the corresponding coordinates.

\begin{figure}[t!]
\label{fig:1}
\begin{center}
\includegraphics[width=250pt,height=270pt]{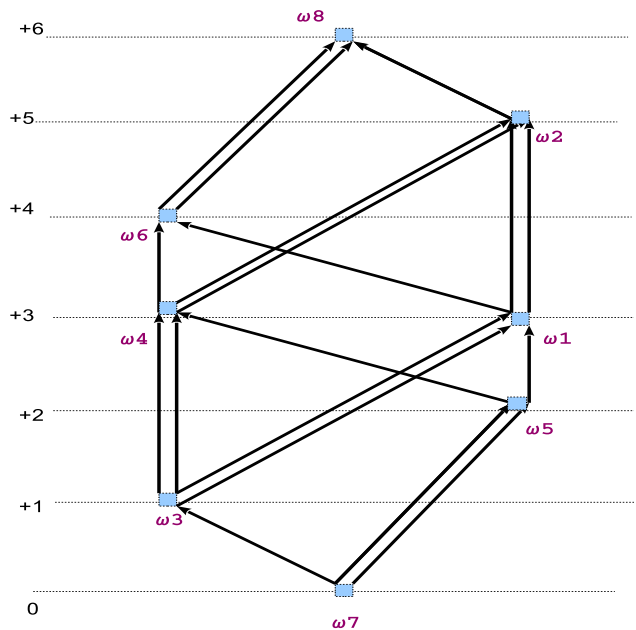}
\caption{1-d and 2-d trajectories connecting two singular points $\omega_{i}$, which are not covered by the trajectories of the higher dimensions.}
\end{center}
\end{figure}

We combine these data to obtain the following diagram, depicting the structure of the trajectories of the system under consideration (see figure \ref{fig:1}): at this diagram we use double arrows to describe those couple of points, which are, on one hand, direct neighbours in the sense that there are no ``broken'' trajectory, connecting them, and on the other hand there exists a two-dimensional family of trajectories going directly from one of them to another.

As one sees, this diagram is combinatorically isomorphic to the Hasse diagram of the $C_2$ root system  which appears, for instance, in the case of the classical group $Sp(4,\mathbb R)$ (see \cite{FS,CSS17}); the difference in the labelling of the vertices is caused by the choice of $\Lambda$ (for other $\Lambda$ we shall have an isomorphic diagram, but with different vertex labels). On the other hand, the appearance of the double arrows is a new phenomenon: in all the situations, we considered before, all the direct neighbours were connected only by a single trajectory (which corresponds to the usual arrow on the diagram). We suppose that this phenomenon is caused by the fact that the root spaces in the non-split case we consider here are not $1$-dimensional any more.

Finally, observe that the points, not directly connected by the arrows on our diagram, but such that there exists a ``broken'' path between them on it, are also connected by trajectories of our system; the dimension o the space, spanned by these trajectories is equal to the sum of the weights of the arrows in the broken trajectory, connecting these points, i.e. every single arrow has weight $1$ and every double arrow has weight $2$. This gives a partial justification for the existence of double arrows: the dimension of our phase space is equal to $6$ (in the case of $Sp(4,\mathbb R)$ it was equal to $4$, see \cite{CSS17}), so, if there were only single arrows, we would lack dimension between the top and the bottom points.

\paragraph{Acknowledgments}
The work of Yu.B. Chernyakov was supported by grant RFBR-15-02-04175. The work of G.I Sharygin was supported by grant RFBR-15-01-05990. The work of A.S. Sorin  was partially supported by the RFBR Grants No. 16-52-12012 -NNIO-a, No. 15-52-05022-Arm-a
and by the DFG Grant LE 838/12-2.


\begin{thebibliography}{99}

\bibitem{CSS14}
Yu.B. Chernyakov, G.I. Sharygin, A.S. Sorin, Bruhat Order in Full Symmetric Toda System, Commun. Math. Phys. 330, 367В–399 (2014),  [arXiv:1212.4803].

\bibitem{CSS15}
Yu.B. Chernyakov, G.I. Sharygin, A.S. Sorin, Bruhat Order in the Full Symmetric $sl_n$ Toda Lattice on partial flag space, SIGMA, 12 (2016), 084,  [arXiv:1412.8116].

\bibitem{CSS17}
Yu.B. Chernyakov, G.I. Sharygin, A.S. Sorin, Phase portraits of the generalized full symmetric Toda systems on rank 2 groups, Theor. Math. Phys. 193 (2017) 2, 1574–1592,  [arXiv:1512.05821].

\bibitem{Hel}
S. Helgason, Differential geometry, Lie group and Symmetric space, Academic Press, 1978.

\bibitem{VinOn}
 E. B. Vinberg, A. L. Onishchik: Lie groups and algebraic groups. Springer, 1990.

\bibitem{deMPe}
F. De Mari, M. Pedroni, Toda flows and real Hessenberg manifolds. J. Geom. Anal., 9 no.4 (1999), 607 -- 625.

\bibitem{BBR}
A. M. Bloch, R. W. Brockett and T. S. Ratiu, Completely Integrable Gradient Flows, Comm. Math. Phys, 147 (1992), 57--74.

\bibitem{BG}
A. M. Bloch and M. Gekhtman, Hamiltonian and gradient structures in the Toda flows, J. Geom. Phys. 27 (1998), 230 -- 248.

\bibitem{FS}
P. Fre, A.S. Sorin, The arrow of time and the Weyl group: all supergravity billiards are integrable, Nucl. Phys., B 815 (2009), 430, [arXiv:0710.1059].

\bibitem{CS}
Yu.~B.~Chernyakov, A.~S.~Sorin,
Explicit Semi-invariants and Integrals of the Full Symmetric $\mathfrak{sl}_n$ Toda Lattice, Lett. Math. Phys. (2014), 104: 1045 -- 1052, [arXiv:1306.1647].

\end{thebibliography}
\end{document}